\documentclass[aps,prl,twocolumn,superscriptaddress,showpacs,floatfix,nofootinbib]{revtex4}

\usepackage{color}
\usepackage{graphicx}

\newcommand{\lsim}{\mathrel{\mathop{\kern 0pt \rlap
      {\raise.2ex\hbox{$<$}}}\lower.9ex\hbox{\kern-.190em $ \sim$}}}
\newcommand{\gsim}{\mathrel{\mathop{\kern 0pt
      \rlap{\raise.2ex\hbox{$>$}}}\lower.9ex\hbox{\kern-.190em $\sim$}}}
\newcommand{\beq}{\begin{equation}}
\newcommand{\eeq}{\end{equation}}

\newcommand{\be}{\begin{equation}}
\newcommand{\ee}{\end{equation}}
\newcommand{\bea}{\begin{eqnarray}}
\newcommand{\eea}{\end{eqnarray}}
\newcommand{\beqarr}{\begin{eqnarray}}
\newcommand{\eeqarr}{\end{eqnarray}}

\newcommand{\lum}{\mathcal{L}}

\begin{document}

\title{A dark matter interpretation for the ARCADE excess?}
\thanks{Preprint number: DFTT 19/2011}

\author{N. Fornengo}
\affiliation{Dipartimento di Fisica Teorica, Universit\`a di Torino, I-10125 Torino, Italy}
\affiliation{Istituto Nazionale di Fisica Nucleare, Sezione di Torino I-10125 Torino, Italy}
\author{R. Lineros}
\affiliation{IFIC, CSIC--Universidad de Valencia, Ed. Institutos, Apdo. Correos 22085, EÐ46071 Valencia, Spain and MultiDark fellow}
\author{M. Regis}
\affiliation{Dipartimento di Fisica Teorica, Universit\`a di Torino, I-10125 Torino, Italy}
\affiliation{Istituto Nazionale di Fisica Nucleare, Sezione di Torino I-10125 Torino, Italy}
\author{M. Taoso}
\affiliation{IFIC, CSIC--Universidad de Valencia, Ed. Institutos, Apdo. Correos 22085, EÐ46071 Valencia, Spain and MultiDark fellow}
%

%

\date{\today}

\begin{abstract}

The ARCADE 2 Collaboration has recently measured an isotropic radio emission which is significantly brighter than the expected contributions from known extra--galactic sources. The simplest explanation of such excess involves a ``new'' population of unresolved sources which become the most numerous at very low (observationally unreached) brightness. We investigate this scenario in terms of synchrotron radiation induced by WIMP annihilations or decays in extragalactic halos.
Intriguingly, for light--mass WIMPs with thermal annihilation cross--section, the level of expected radio emission matches the ARCADE observations.

\end{abstract}


\maketitle


The detection of a non-gravitational signal of Dark Matter (DM) would be one of the greatest pillar of modern physics, simultaneously confirming our views of cosmology, astrophysics and particle physics. This possibility might be not far ahead if DM is in the form of Weakly Interacting Massive Particles (WIMPs), which currently are the most investigated class of DM candidates
 \cite{Jungman:1995df}. Signatures of this scenario include a multi--wavelength spectrum associated to radiative emissions involving electrons and positrons generated in WIMP annihilations or decays (for a recent review on this topic, see {\em e.g.}, Ref. \cite{Profumo:2010ya}).

Recently, the balloon--borne experiment ARCADE 2 (Absolute Radiometer for Cosmology, Astrophysics and Diffuse Emission)~\cite{Singal:2009xq} reported radio measurements of the sky temperature at frequencies ranging from 3 to 90 GHz. Observations have been performed on a region which is roughly an annulus centered at (l, b) = (70, 0) with radius and widths of 30 and 20 degrees, respectively \cite{Kogut:2009xv}.

An isotropic component can be isolated from the ARCADE data by subtracting foreground Galactic emission~\cite{Fixsen:2009xn}. Surprisingly, the level of the remaining flux (which has been interpreted in terms of extra--galactic sky temperature) is about 5--6 times larger than the total contribution from the extra--galactic radio sources detected in current surveys~\cite{Seiffert:2009xs,Gervasi:2008rr}. Even extrapolating the source number counts to lower (unreached) brightness, such excess still remains.

Most sources of systematic effects which could explain the ARCADE excess
have been ruled out~\cite{Fixsen:2009xn}.
An astrophysical galactic origin appears to be rather unlikely (see discussions in Refs. \cite{Kogut:2009xv} and \cite{Singal:2009dv}). Indeed, free--free emission has been
excluded based on the spectral shape, and diffuse Galactic synchrotron foreground is estimated using two different methods (namely, a co--secant dependence on Galactic latitude and the correlation between radio and atomic line emissions), which agree well among each other.

The observed isotropic temperature can be fitted by the CMB blackbody contribution plus a power law:

\beq
T(\nu) = T_0 \,\frac{h\nu/(k\,T_0)}{\exp[h\nu/(k\,T_0)]-1}+T_s \left(\frac{\nu}{{\rm GHz}}\right)^{\alpha}
\label{eq:temp}
\eeq
where $T_0=2.729\pm0.004$ K~\cite{Fixsen:2009xn} is the CMB thermodynamic temperature. Performing analogous analyses on past surveys at 22, 45, 408, and 1420 MHz similar results are obtained, and fitting all data simultaneously the ARCADE Collaboration derived $\alpha=-2.62\pm0.04$ and $T_s=1.19\pm 0.14$ K~\cite{Fixsen:2009xn}.

Such level of cosmic radio background does not have an immediate explanation in standard astrophysical scenarios. In Ref.~\cite{Singal:2009dv}, radio supernovae, radio quiet quasars and diffuse emission from intergalactic medium and clusters (as well as a missed flux from 
well--known sources) have been considered, concluding that none of them can significantly contribute. A new population of numerous and faint radio sources (able to dominate source counts around $\mu$Jy flux) has to be introduced~\cite{Singal:2009dv,Vernstrom:2011xt}. 
Ordinary star--forming galaxies with a radio to far--infrared flux ratio which increases significantly with redshift can in principle offer a solution.
On the other hand, this possibility is strongly constrained by multi--wavelength observations.
Indeed, the radio to far--infrared emission has to be increased by a factor of 5 above what is observed in local galaxies
\footnote{Radio synchrotron sources are associated to young stars and therefore tightly correlated to the star formation rate of galaxies. Empirically, a nearly linear far IR--radio correlation is reasonably well established (at least for high--brightness and low redshift), see {\em e.g.} the review in Ref. \cite{DeZotti:2009an}.}, while current measurements show very mild evolution, at least up to $z\sim2-3$~\cite{Ponente:2011se}.
An explanation of the ARCADE excess through radiative emission of secondary electrons in 
star--forming galaxies would overproduce the gamma--ray background from pion decays \cite{Lacki:2010uz}. The same is true also for primary electrons unless such putative galaxies have extremely low gas density (and, in turn, low ratio of primary electrons to pions) or extremely efficient proton escape.
 The picture that seems to emerge from ARCADE measurements~\cite{Fixsen:2009xn,Singal:2009dv} and subsequent interpretations~\cite{Singal:2009dv,Vernstrom:2011xt,Ponente:2011se,Lacki:2010uz} suggests the need for a population of numerous and faint synchrotron sources generated by primary electrons with a hard spectrum and with no or very faint correlated mechanisms at infrared and gamma--ray frequencies.

In our current understanding of structure clustering, any luminous source is embedded in a DM halo, and therefore extragalactic DM halos can be seen as the most numerous source population. The flux induced by WIMP annihilations/decays is predicted to be very faint. It is associated to primary electrons and positrons generated as final state of annihilation/decay, and WIMP models with large annihilation/decay branching ratios into leptons induce hard spectra of $e^+/e^-$ with very faint gamma--ray counterpart (and, of course, no straightforward thermal emission). Therefore, WIMP sources represent an ideal candidate to fit the ARCADE excess and in this Letter we quantitatively investigate such possibility.




\begin{figure}[t]
\includegraphics[width=0.4\textwidth]{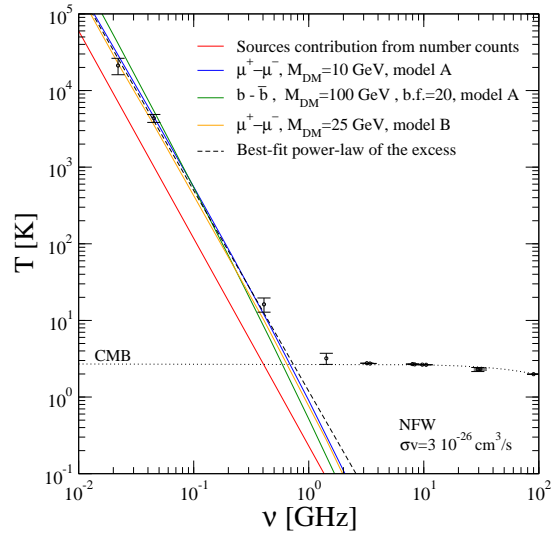}
\caption{Extragalactic radio background as derived by ARCADE~\cite{Fixsen:2009xn}, together with 
three possible interpretations of the low--frequency ($<10$ GHz) excess in terms of WIMP annihilations (blue, green, and orange curves, see text for details). 
The astrophysical source contribution estimated from number counts (red line), the CMB contribution (black--dotted line), and a best--fit power--law of the excess (black--dashed line) are also reported ~\cite{Seiffert:2009xs}.
}
\label{fig:totI}
\end{figure}

Assuming a one-to-one relation between the mass $M$ of extragalactic DM halos  and the intrinsic luminosity $\lum$ of the source, the total isotropic intensity per solid angle at a given frequency $\nu$ is given by (for a more detailed derivation of equations considered in the following, see, {\rm e.g.} Refs. \cite{Zhang:2008rs,noi}):
\be
\nu\,I_\nu=\frac{c\,\nu}{4\pi}\int \frac{dz}{(1+z)\,H(z)}\int_{M_c} dM \frac{dn}{dM}(M,z)\,\lum(E,z,M)\;, 
\label{eq:intgen1}
\ee
where $z$ is the redshift, $H$ is the Hubble rate, $M_c$ is the minimum mass of an emitting halo, 
and the luminosity $\lum$ is function of the redhsifted energy $E=E_\nu(1+z)$ with $E_\nu=h\nu$.
The luminosity function, including also the contribution of substructures within the DM halo, can be written as:
\beq
\lum = (1-f)^a\,\int_0^{R_v} d^3r \frac{d\hat N_i}{dE}+\int dM_s\frac{dn_s}{dM_s}\int_0^{R_v} d^3r \frac{d\hat N_i}{dE}
\label{eq:intgen2}
\eeq
where $R_v$ is the virial radius of the DM profile $\rho$ and $f$ is the fraction of halo mass in substructures (with $a=1$ and $a=2$ for decaying and annihilating DM, respectively).
In Eqs. (\ref{eq:intgen1}) and (\ref{eq:intgen2}), $dn/dM(M)$ and $dn_s/dM_s(f,M_s,M)$ denote the mass function of the DM halo and of substructures, respectively.
For synchrotron emission:
\beq
\frac{d\hat N_i}{dE_\nu}=2\,\int^{M_{\chi}}_{m_e}dE'\, P_{\rm syn}(\nu,B,E')\cdot n_e
\eeq
where $m_e$ is the electron mass, $P_{\rm syn}$ is the synchrotron power~\cite{Rybicki}, $B$ denotes the magnetic field, and $\nu$ is the frequency of emission (as opposed to frequency of observation in Eq.(\ref{eq:intgen1})).
The electron/positron equilibrium number density $n_e$ is obtained solving a transport equation for $e^-/e^+$ injected by DM with an energy spectrum set by $dN_e/dE_e$ \cite{noi}. 
The source terms of this equation for annihilating and decaying DM are:
\beq
Q_a=\frac{(\sigma_a v)}{2\,M_{\chi}^2}\,\rho^2\,\frac{dN_e}{dE_e}\;\;\;,\;\;\;Q_d=\frac{\rho}{\tau_d\,M_{\chi}}\,\frac{d N_e}{dE_e} \, ,
\eeq
where $M_{\chi}$ is the mass of the DM particle, $(\sigma_a v)$ is the non--relativistic
annihilation cross section and $\tau_d$ the decay rate.

\begin{figure}[t]
\includegraphics[width=0.4\textwidth]{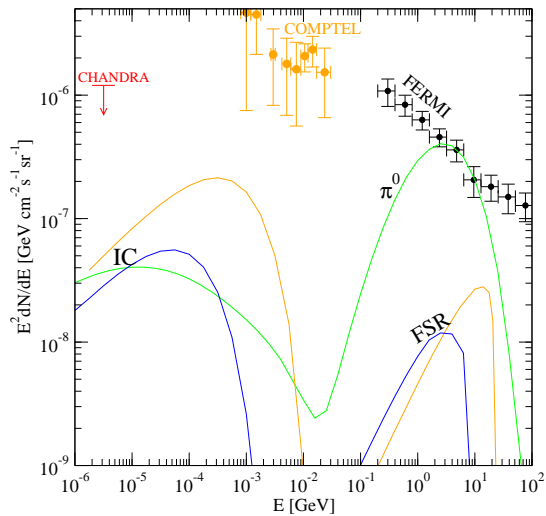}
\caption{X--ray and gamma--ray fluxes for the three benchmark WIMP cases shown in Fig. \ref{fig:totI}. 
The CHANDRA~\cite{Hickox:2007gj} bound in the X--ray band and the COMPTEL~\cite{COMPTEL} and FERMI~\cite{Abdo:2010nz} extragalactic gamma--ray
fluxes are shown.}
\label{fig:totII}
\end{figure} 

For what concerns the `astrophysical' parameters, we focus on two benchmark cases, which are fairly realistic (for a detailed discussion of impact of astrophysical uncertainties, see Ref. \cite{noi}). 
We adopt the halo mass function $dn/dM$ from Ref. \cite{Sheth:1999mn}, recent N--body simulation results for concentration of halos \cite{MunozCuartas:2010ig}, a DM distribution inside halos following a NFW profile \cite{Navarro:1996gj}, and the minimum halo mass is set to $M_{c}= 10^6 M_{\odot}$ (model A, in order to consider only objects for which we can guess a reasonably large magnetic field), and to $M_{c}= 10^{-6} M_{\odot}$ (model B). The contribution from substructures is modeled such that $f=10$\% of the total mass is in substructures and $dn_s/d\ln(M_s)\propto 1/M_s$, which leads to a boost in the signal of $b.f._{\rm sub}\simeq7$ (model A) and no boost in model B. Magnetic field is assumed to be constant in space and time with magnitude $B=10\mu G$ (model A) and $B=2\mu G$ (model B), and $e^+/e^-$ are assumed to radiate at the same place where they are injected.\footnote{Notice that for the DM candidate of specific interest here ({\em i.e.} light or inducing a soft spectrum of $e^+/e^-$), electrons are mostly emitted at GeV energies, namely are injected at energies relevant for radiation at GHz frequencies. Therefore they don't travel significant distances before radiating, while electrons emitted at larger energies would take some time to cool down, and diffusion and escape time would become much more relevant.}
The normalization of the emission roughly decreases by an order of magnitude going from $B=10\,\mu G$ to $B=1\,\mu G$, increases by two orders of magnitude going from $M_c= 10^6 M_{\odot}$ to $M_c= 10^{-6} M_{\odot}$, scales linearly with $b.f._{\rm sub}$, and is mildly dependent on halo mass function, concentration and profile of DM.\footnote{For example, the case of an isothermal cored profile leads to a reduction in the intensity by, roughly, a factor of 1.5.}

The excess spectrum reported by ARCADE and described in Eq. (\ref{eq:temp}) is rather hard, and requires a hard electron/positron spectrum $d N_e/dE_e$. This can be produced by DM scenarios with a large branching ratio of annihilation/decay into leptons. For illustrative purposes, we chose the $\mu^+-\mu^-$ channel. 
To reproduce the absolute normalization of the excess with a `thermal' annihilation rate $ (\sigma_a v) =3\cdot 10^{-26}{\rm cm^3s^{-1}}$ in our benchmark model A, we need a WIMP with $M_{\chi}=10$~GeV ($M_{\chi}=25$~GeV in model B), which, even though induces a slightly softer spectrum than the 
best--fit power--law, provides a reasonable agreement with the data, as shown in Fig.~\ref{fig:totI}. For this benchmark case we have $\chi^2/{\rm dof}=26.9/13$.
The actual best--fit for model A is obtained with a mass $M_{\chi}\sim30$~GeV, ($\chi^2/{\rm dof}=14.3/13$), but at the price of increasing the cross section by one order of magnitude. 
The fact that light DM, in the 10 GeV mass range, can fairly well reproduce the ARCADE excess, without the need of unrealistically large DM overdensities
is particularly interesting, especially in light of recent claims of signals compatible with a DM interpretation from direct detection experiments 
(DAMA \cite{dama2010}, CoGeNT \cite{cogent}, and CRESST \cite{CRESST}), that can be in fact accommodated with a $\sim 10$ GeV WIMP \cite{DDth}. 
In the case of the ARCADE excess, the best option to explain the effect in terms of
DM annihilation requires a light DM particle which annihilates mainly into leptons, and therefore that does not couple dominantly to quarks (coupling relevant to the
direct detection scattering cross--section).
Nevertheless, it is not very difficult, from the model--building point of view, to foresee a model where a DM candidate, which annihilates mainly into leptons, still has relatively large scattering cross-section off nuclei. For a concrete example, see {\em e.g.}, Ref. \cite{Boucenna:2011hy}.
Note also that the radio emission in the Milky-Way halo induced by WIMPs fitting ARCADE data can either easily satisfy constraints (for cored galactic DM profiles) or be close to a possible detection in the central region of our Galaxy (for cuspy profiles)~\cite{radiopaper}.

Similar conclusions on the viability of a light `leptophilic' DM particle in explaining the
ARCADE data can be also drawn in the decaying case: for a DM mass of $M_{\chi}=10$ GeV,
the excess is reproduced if the lifetime is $\tau_d=3\cdot10^{27}$ s (with a curve similar to the one shown for the annihilating case).

Since the ARCADE excess can be explained by DM annihilation/decay in terms of
sizable production of electrons and positrons, emissions of X--rays and gamma--rays 
by means of inverse--Compton processes on interstellar radiation fields (here we include CMB only) and direct
production of gamma--rays from the DM particle annihilation (either by production
of neutral pions or by Final--State--Radiation (FSR)) are present and have to be checked against
available bounds. For the benchmark cases considered, these multi--wavelength constraints are easily satisfied, as shown 
for X- and $\gamma$-rays in Fig. \ref{fig:totII}. 

In Fig.~\ref{fig:totI}, we show also the case of a more `classic' WIMP candidate with 100 GeV mass and hadronic annihilation channel ($\bar b b$ pair in the shown benchmark). This scenario is less appealing than previous cases. The spectrum is relatively too soft in order to reproduce well the ARCADE data and the fit is worse ($\chi^2/{\rm dof}=49.5/13$) than for DM annihilating into leptons; note that the excess is sizable up to at least 3 GHz (although not clearly visible in Fig. \ref{fig:totI} due the smallness of the error bars and the scale of the plot), so a viable explanation has to roughly overlap to the dashed best-fit curve up to those frequencies. Moreover, since now the DM mass is larger, the required boost factor is accordingly larger (by a factor of 20 in this specific case),
which can stem from a larger annihilation cross section related to a non standard formation of DM relic density, from a Sommerferld enhancement, or from a larger contribution of substructures with respect to what considered here (see {\em e.g.}, \cite{boost} for further details on possible boost factors).
Heavier DM with hadronic annihilation/decay final states is also more strongly constrained by the $\gamma$--ray channel, as can be seen in Fig. \ref{fig:totII}.

\begin{figure}[t]
\includegraphics[width=0.4\textwidth]{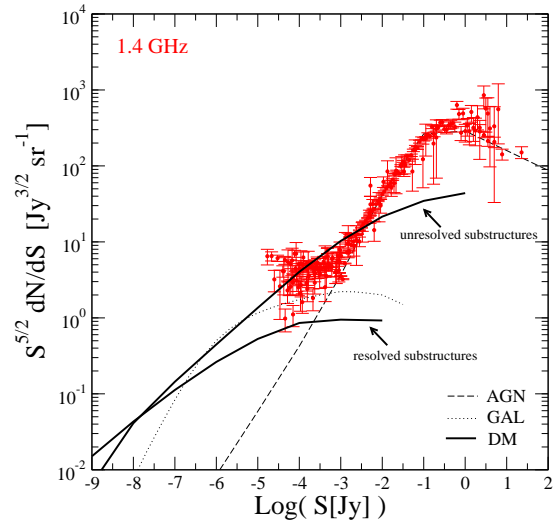}
\caption{Differential number counts for AGNs (dashed line), star--forming galaxies (dotted line), and the same 10 GeV benchmark DM model (solid lines) shown in Figs. \ref{fig:totI} and \ref{fig:totII}. For DM, we consider a case such that all substructures are resolved, and an opposite case where all substructures are unresolved. For data and astrophysical models, see \cite{DeZotti:2009an} and references therein.}
\label{fig:counts}
\end{figure}

As a further analysis on the radio emission arising from light DM annihilation/decay,
able to adapt to the ARCADE excess, we show in Fig. \ref{fig:counts}
the differential number counts of sources at 1.4 GHz. 
If we assume all substructures to be unresolved, they mainly boost the signal of large and bright halos (since the latter host more subhalos). On the contrary, if all substructures are assumed to be resolved, counts drop much more slowly at low brightness. To bracket uncertainties related to the possibility of resolving substructures in the future, these two extreme cases are shown
in Fig. \ref{fig:counts} as solid lines, for the same 10 GeV DM benchmark of Fig. \ref{fig:totI}. As discussed above, the key point for our analysis is that in both scenarios the number of DM sources definitely becomes dominant over astrophysical contributions
(AGN, star--forming galaxies) at the sub-$\mu$Jy level. The contribution of star--forming galaxies, which is dominant over AGN emission at low fluxes, decreases more rapidly (assuming FIR/radio correlation holds at all redshift), than the expected contribution from DM, both in the case
of resolved and unresolved substructures. From Fig.~\ref{fig:counts} we notice that
the flattening at low brightness in current data, although it can be easily accounted for by standard astrophysical populations (for a review, see {\em e.g.} Ref.~\cite{DeZotti:2009an}),
nevertheless could be well fitted by a DM model between the two extreme cases presented in Fig.~\ref{fig:counts}.

In summary, we discussed the possibility that synchrotron emission induced by WIMP annihilations can account for the isotropic radio component measured by the ARCADE 2 Collaboration.
Although galactic or extragalactic astrophysical interpretations of the excess cannot be excluded, they currently present some puzzling issues \cite{Kogut:2009xv,Seiffert:2009xs,Singal:2009dv}.
Under reasonable assumptions for clustering, we found that light-mass WIMPs producing hard-spectrum electrons and positrons (as in the case of leptonic annihilation channels) in extragalactic halos with a `thermal' annihilation rate can fit the excess and satisfy constraints at other wavelengths.
A population of sources which can generally explain ARCADE measurements has to become the most numerous at brightness around $\mu$Jy, so it will be certainly studied in details by SKA \cite{SKA}, and possibly also by its precursors, ASKAP \cite{ASKAP} and MeerKAT \cite{MeerKAT}.
If the excess is due to extragalactic DM, a clear discovery of a non--gravitational signal of DM might be not far ahead.
A dedicated study of closest and brightest (in terms of DM--induced signal) objects with current radio telescope (e.g., ATCA \cite{ATCA} and EVLA \cite{EVLA}) can start to probe this scenario in the near future.

\acknowledgments

We acknowledge research grants funded jointly by Ministero
dell'Istruzione, dell'Universit\`a e della Ricerca (MIUR), by
Universit\`a di Torino and by Istituto Nazionale di Fisica Nucleare
within the {\sl Astroparticle Physics Project} (MIUR contract number: PRIN 2008NR3EBK;
INFN grant code: FA51). RL and MT were supported by the EC contract UNILHC
PITN-GA-2009-237920, by the Spanish grants FPA2008-00319 and MultiDark
CSD2009-00064 (MICINN), and PROMETEO/2009/091 (Generalitat Valenciana).
NF acknowledges support of the spanish MICINN
Consolider Ingenio 2010 Programme under grant MULTIDARK CSD2009- 00064.
This work was partly completed at the Theory Division of CERN in the context of the TH--Institute `Dark Matter Underground and in the Heavens' (DMUH11, 18-29 July 2011).

\end{document}